\documentclass[amsmath,amssymb,prl,twocolumn]{revtex4}
\usepackage{graphicx}
\usepackage{dcolumn}
\usepackage{bm}
\begin{document}

\newcommand{\be}{\begin{equation}}
\newcommand{\ee}{\end{equation}}
\newcommand{\ber}{\begin{eqnarray}}
\newcommand{\eer}{\end{eqnarray}}


\title{Symmetry considerations and development of pinwheels in visual maps}

\author{Ha Youn Lee, Mehdi Yahyanejad, and Mehran Kardar} 
\affiliation{ 
Department of Physics, 
Massachusetts Institute of Technology, Cambridge, Massachusetts 02139}

\begin{abstract}
\indent  
Neurons in the visual cortex respond best to rod-like stimuli of given 
orientation.  
While the preferred orientation varies continuously across most of the cortex, 
there are prominent {\em pinwheel} centers around which all orientations are 
present.   
Oriented segments abound in natural images, and tend to be {\em collinear};
neurons are also more likely to be connected if their preferred orientations
are aligned to their topographic separation.
These are indications of a {\em reduced symmetry} requiring joint rotations
of both orientation preference and the underlying topography.
We verify that this requirement extends to cortical maps of monkey and cat
by direct statistical analysis.
Furthermore, analytical arguments and numerical studies  
indicate that pinwheels 
are generically stable in evolving field models which couple orientation and topography.
\end{abstract}

\maketitle

\section{Introduction}\label{Introduction}
The preferential response of cells in the primary visual cortex 
to lines of a particular orientation has been 
known for over forty years\cite{hubel:evolution},
yet remains a subject of intense experimental study and modelling.
Early models were simple structural 
arrangements of local iso-orientation columns
into regular arrays\cite{hubel:structure,Obermayer90,Braitenberg79}.
Intricate maps of global patterns of orientation preference over the cortex,
obtained by optical imaging\cite{blasdel86,blasdel92},
revealed more complex arrangements.
Thus, later models focused on the development of orientation preference (OP)
in networks of neurons whose connectivity is modified in response to
stimuli\cite{Kohonen,Linsker,Malsburg}.
Obtaining large scale patterns of OP with many pinwheels is computationally 
costly with the latter models\cite{NeuralNet};
drastically simplified models generate large static maps essentially from
bandpass filtered white noise\cite{Rojer,Niebur,Grossberg}.

Analytical understanding of the development of visual maps, and its
connections to other problems in pattern formation, is best obtained 
in terms of {\em evolving fields}.  
In this framework, OP is modelled by a 
director field ${\bf s}\equiv\left(s_x(x,y), s_y(x,y)\right)$, indicating the preferred 
orientation at location ${\bf r}\equiv(x,y)$ on the cortex.  
The field ${\bf s}\left({\bf r},t\right)$ then evolves in time according to some 
development rule that depends on its configurations at 
earlier times\cite{Swindale82,Swindale92}.  
Wolf and Geisel (WG) have shown\cite{wolf-nature} that a large number of such evolutions
can be summarized through a dynamical equation 
$\partial_t {\bf s}\left({\bf r},t\right)=F\left[{\bf s}\right]$.  
(WG combine the two components into a single complex field $z=\left(s_x+is_y\right)^2$.)
Common elements in models of evolving fields are: 

\par\noindent{\em (a)} Starting from an initial condition with little OP, 
there is a rapid onset of selectivity governed by $L\left[{\bf s}\right]$,
the {\em linear} part of the functional $F\left[{\bf s}\right]$.  
The characteristic length scale observed in cortical maps is implemented by
a linear operator that causes  maximal growth of features of wavelength $\Lambda$,
{\em i.e.} acting as a  `band-pass filter' in the parlance of circuits.
It is possible to follow the linear development analytically:
WG show that the density of pinwheels 
(zeros of the field $z({\bf r})$) has to be larger than $\pi/\Lambda^2$
in this regime.

\par\noindent{\em (b)} Because the linear evolution leads to unbounded growth of OP, 
nonlinearities are essential for a proper saturation of the field.  
Although analytical studies of nonlinear development are difficult, 
numerical simulations indicate that the OP patterns continue to change
(albeit more slowly) even after their magnitudes have saturated.  
More importantly, the pinwheels typically annihilate in pairs,
giving way to a rainbow pattern of wavelength $\Lambda$.  
To maintain pinwheels, development has to be stopped, or extrinsic elements such
as inhomogeneities that trap the pinwheels have to be introduced\cite{Fnoise}.
Because the neural processes that lead to OP are still not fully understood,
the stability of pinwheels has not been a topic of much study amongst neuroscientists.
Nevertheless, the search for intrinsically stable pinwheel patterns has motivated some 
recent studies\cite{KC,Wthesis}.
We propose here an alternative explanation, 
demonstrating that evolving field models with proper rotational symmetry
generically lead to patterns with stable pinwheels.

\begin{figure}
\centering
\includegraphics[height=5cm,width=5cm]{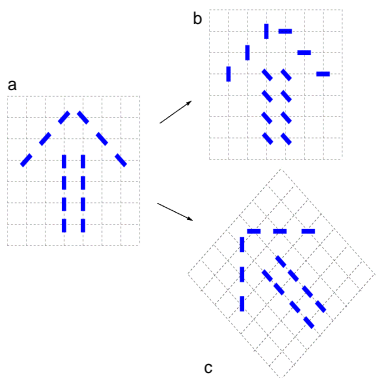}
\caption[]{ {\bf a} depicts the image of an arrow formed by oriented solid lines, 
on a topographic grid of dotted lines.
In {\bf b}, each solid line is rotated anti-clockwise by 45$^o$ independent of the grid.
The thus `rotated' image bears little resemblance to the original.
In {\bf c}, there is simultaneous rotation of the grid and the solid lines, as the whole
image is rotated.}
\label{fig:rotation}
\end{figure}

Symmetry considerations are paramount in problems of pattern formation.
Because all directions are more or less equally present in cortical maps,
practically all models of OP (certainly those summarized in WG) 
assume that different orientations are equivalent\cite{Fanisotropy}.
The full rotational symmetry is implemented by requiring the evolution of 
${\bf s}\left({\bf r},t\right)$ to be unchanged if all angles are rotated together.
This rotation is independent of the topographic space ${\bf r}$, 
which is also assumed to be isotropic (no preferred directions).  
Two versions of rotation are illustrated in Fig.~\ref{fig:rotation}. 
Figure \ref{fig:rotation}b displays a collection of oriented lines that are rotated
independently of the background grid from Fig.~\ref{fig:rotation}a.  
We propose that the appropriate symmetry for OP maps is simultaneous 
rotations of the orientations 
{\em and} the underlying space, as illustrated in Fig.~\ref{fig:rotation}c.  

 The observational evidence for the reduced symmetry is reviewed
in Sec.~\ref{observations}.
As suggested by  Fig.~\ref{fig:rotation}, the absence of full rotation symmetry
in natural images is expected, and in fact demonstrated in Ref.~\cite{Gilbert}.
There is also an evidence that neural connectivities are preferentially
linked along the axis of OP\cite{bio:shrew}. 
We present a statistical analysis of OP maps from monkey and cat,
which also supports the lack of full rotation symmetry.
Consequences of reduced symmetry in evolving field models are
discussed in Sec.~\ref{models}.
A linear analysis indicates that the reduced symmetry introduces an additional
time scale into the problem, and an interval in which the pinwheel density
can actually increase by pair creations.
Vectorial versions of {\em center--surround} interactions are then used 
in numerical simulations of model with joint rotation symmetry.
The simulations result in patterns with intrinsically stable pinwheels,
and histograms of OP similar to those obtained from cat and monkey maps.

\section{Observational evidence of reduced symmetry in cortical maps}\label{observations}

Casual consideration of scenes strongly suggests that the persistence of edges 
of stationary objects (as in Fig.~\ref{fig:rotation}), or of tracks of moving ones, 
leads to oriented segments that cannot be rotated independent of their background.
This expectation has been confirmed and quantified by statistical tests in Ref.~\cite{Gilbert},
where an orientation was assigned to each pixel of images from the natural world.
The primary query of Ref.~\cite{Gilbert} was the range 
and directionality of correlations in orientation.
They observed that correlations depend on the relative angles in the 
topographic space, in a manner consistent with a collection of circles.

Since the task of the visual system is to extract information from observed images,
it is likely that the neural connections that carry out the associated computations
are influenced by symmetries and anisotropies of the natural scenes.
Contemplation of the Hebbian rule\cite{Hebb} ``neurons that fire together
wire together," suggest that there should be more connections between
neurons whose shared OP is collinear to their topographic separation.
Indeed, biocytin injections which map the `horizontal' connections of neurons
have been combined with optical imaging of the primary visual cortex of the
tree shrew\cite{bio:shrew}.
The connections from an injection site are anisotropic, preferentially extended
along the axis of OP at the site.
Although less pronounced, similar anisotropies are also observed in maps from
monkey\cite{bio:monkey} and cat.
Such connectivities are incompatible with rotations of OP independent
of the underlying topography.
A map with all OPs rotated by a fixed angle would require a  different set
of horizontal connections.
 
\begin{figure}
\centering
\includegraphics[height=2cm,width=3cm]{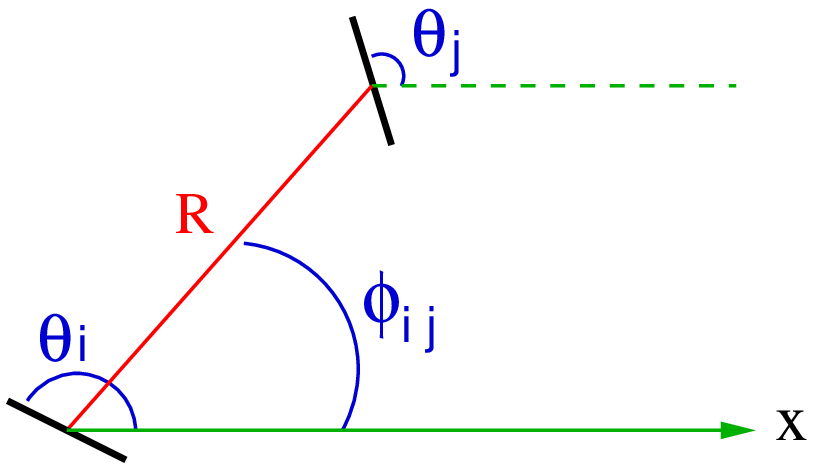}\\a\\
\includegraphics[height=5cm,width=6cm]{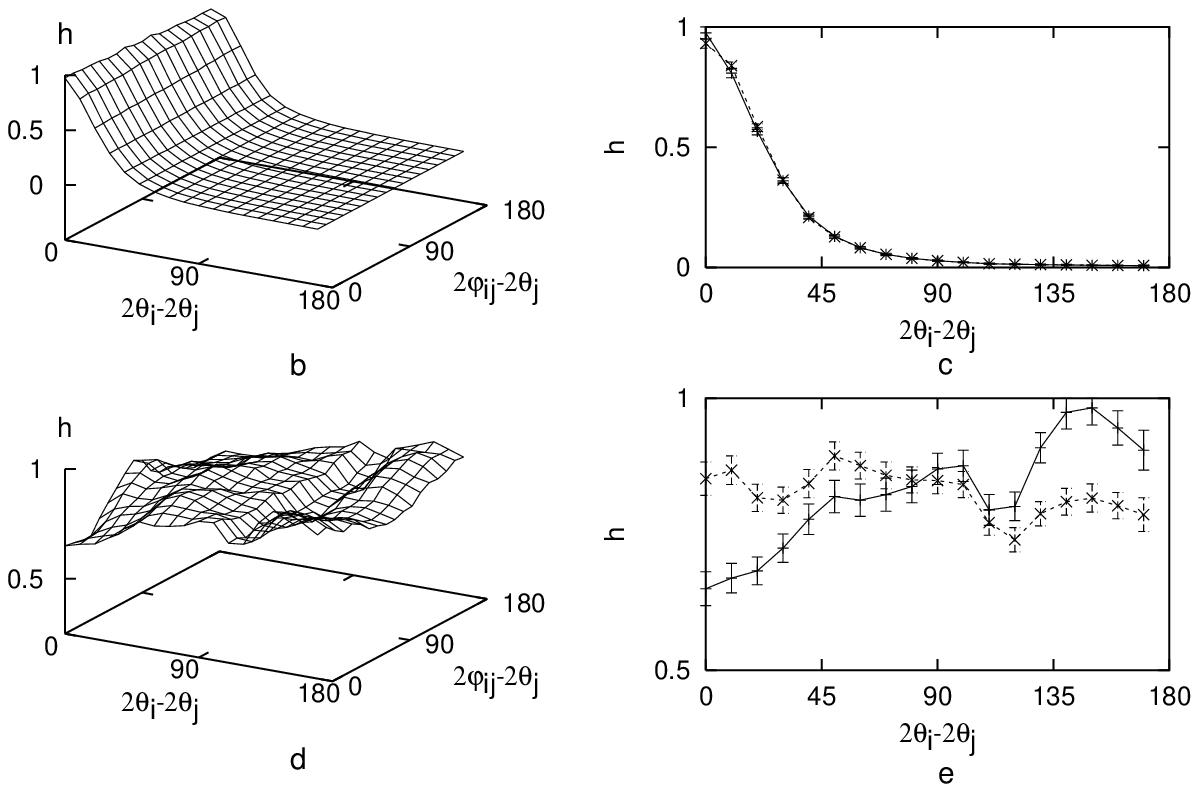}\\
\caption[]{
Histograms of OP from a cortical map of monkey. 
a. The relative orientation $2\left(\theta_i-\theta_j\right)$, between two pixels $i$ and
$j$ at a distance $R$, is one argument of the histogram; 
the second is the OP of one point measured relative
to the line joining the two pixels (at angle $\phi_{ij}$).
Full histograms are shown on the left column, while the right column is for
$2\left(\phi_{ij}-\theta_j\right)=0^0$ (solid line) or $90^o$ (dotted line).
{\bf b} and {\bf c} are for short separations of 5 to 10 pixel spacings, and show
no dependence on the relative angle.
By contrast, there is a small but clear indication of a coupling to the underlying topography
in {\bf d} and {\bf e} which are taken at distances of 70 to 75 pixels, comparable
to the separations of pinwheels. 
Such dependence indicates the lack of full rotation symmetry in the map.
}
\label{fig:monkeymap}
\end{figure}

To test the hypothesis that cortical maps of OP also reflect the reduced 
rotation symmetry, we undertook statistical tests of a map of monkey 
(in the form of 360$\times$480 pixels,  provided by K. Obermeyer, Technical University of Berlin, Berlin).
At each point $i$ of the map, there is an orientation angle $\theta_i$,
measured relative to an arbitrary axis;
two points $i$ and $j$, separated by a distance $R$ form an angle
$\phi_{ij}$ with the same axis, as indicated in Fig.~\ref{fig:monkeymap}a.
Binning into intervals of $10^o$, we make joint histograms of the form 
$h_R\left[2\left(\theta_i-\theta_j\right), 2\left(\phi_{ij}-\theta_j\right)\right]$.
(The factor of two is introduced since the orientation is defined from 0 to $\pi$.)
The second argument measures the angle relative to the line joining points $i$ and $j$. 
If the orientations are independent of topography,
the histograms should be independent of their second argument.
This is not the case for the monkey histograms shown on the left column in Fig.~\ref{fig:monkeymap};
the right column shows cross-sections at $2\left(\phi_{ij}-\theta_j\right)=0^o$ 
and $90^o$ which display maximal contrast for parallel orientations.
The larger probability for $2\left(\phi_{ij}-\theta_j\right)=90^o$ does not violate
expectations based on collinear orientations. 
This is because we do not know the actual topographic axis in our monkey map.
The choice of an arbitrary axis does not modify $\theta_i-\theta_j$, but
shifts the histograms along $2\left(\phi_{ij}-\theta_j\right)$.
The advantage of our method is the ability to detect lack of full rotation symmetry
in the absence of knowledge of topographic axis;
but the lack of this information prevents making a connection to correlations
in visual inputs.

 Figures ~\ref{fig:monkeymap}b and ~\ref{fig:monkeymap}c 
are at separations $R$ which are a fraction of the typical distance between pinwheels,
and show no indication of any dependence on topography.
By contrast, Figs.~\ref{fig:monkeymap}d and ~\ref{fig:monkeymap}e
correspond to values of $R$ comparable to pinwheel separations.
There is now a small, but distinct dependence on the orientation of the line between 
two points; indicating that the OPs do not follow a distribution with full rotational symmetry.
Similar results were obtained for maps from cat (204$\times$372 pixels, provided
by M. Sur and J. Schummers, Massachusetts Institute of Technology, Cambridge), and are 
available as Fig.~\ref{fig:cathisto} in the supporting information.
In both cases the dependence on the second argument is small
(at most around 20\%), and some assessments of its statistical significance is needed.
Since we had access to only one map in each case we made an indirect 
estimate of statistical error by constructing an artificial ensemble of 2000 histograms
though random samplings of 2.9\% of total pixels in the monkey map.
(As described in the supporting information  we tested this sampling procedure
on maps generated by numerical simulations.)
From the thus included errors bars in Fig.~\ref{fig:monkeymap}e, we
conclude that the differences fall outside statistical errors.

\section{Modeling Joint Rotation Symmetry}\label{models}
We believe that the restriction to joint rotation symmetry is an essential aspect of
the OP maps, and should be incorporated into models and analytical studies.
In computational models with neural networks\cite{NeuralNet} 
this is naturally achieved through the choice of proper training set of images. 
How should this be implemented in analytical models of evolving fields?
If the inputs to locations (such as $i$ and $j$ in Fig.~\ref{fig:monkeymap}) are
predominantly parallel, a Hebbian interaction between them would evolve to
minimize $\theta_i-\theta_j$. 
If the OP at $i$ is indicated by a vector ${\bf s}_i$, this interaction can be written as
$J(R)~{\bf s}_i \cdot {\bf s}_j$\cite{Fdouble}.   
Such an interaction, however, makes no reference to the relative orientation
$\theta_j-\phi_{ij}$ and thus cannot represent a response to a preponderance of
inputs that are collinear with the topographic (unit) vector $\hat{\bf r}_{ij}$.
To account for the latter, we could have distinct interactions between components
of ${\bf s}_i$ and ${\bf s}_j$ that are parallel or perpendicular to $\hat{\bf r}_{ij}$;
the difference between them can be represented by a new interaction
of the form $K(R)~\left({\bf s}_i \cdot \hat{\bf r}_{ij}\right)
\left({\bf s}_j \cdot \hat{\bf r}_{ij}\right)$\cite{Fdouble}.
With this distinction, when ${{\bf s}_i}$ and ${{\bf s}_j}$ 
are parallel (perpendicular) to ${\hat r}_{ij}$, 
the strength of interaction is $J(R)+K(R)$ $(J(R))$.

As a specific model, let us assume a set of 
${\bf s}_i(t)$, stimulated by inputs ${\bf p}_i(t)$,
and interactions between 
them that reflect the average activity of ${\bf s}_i(t)$
over previous times. 
The joint activity of ${{\bf s}_i}$ and ${{\bf s}_j}$ contributions
can be decomposed into two components;
${\bf s}_i$ and ${\bf s}_j$ that are parallel to $\hat{\bf r}$
or perpendicular to it.
Both cases contribute to the isotropic interaction
$2J_{ij}(t)=\left[{\bf s}_i\cdot {\bf s}_j\right]_{\rm av.}$,
while the component parallel to $\hat{\bf r}$ give rise to the interaction
$J_{ij}(t)+K_{ij}(t)=\left[\left({\bf s}_i \cdot \hat{\bf r}_{ij}\right)
\left({\bf s}_j \cdot \hat{\bf r}_{ij}\right)\right]_{\rm av.}$.
In the initial stages, the couplings are small and ${\bf s}_i(t)$ merely
follow the inputs ${\bf p}_i(t)$. 
The couplings then evolve to reflect the statistics of inputs:
A tendency for the ${\bf p}_i(t)$ and ${\bf p}_j(t)$ to be parallel leads to
a positive $J_{ij}$, while {\em if and only if} these 
inputs also tend to be collinear, 
a finite $K_{ij}$ is generated. 
Note that if $K_{ij}=0$, we have $\left[{\bf s}_i\cdot {\bf s}_j\right]_{\rm av.}
=2 \left[\left({\bf s}_i \cdot \hat{\bf r}_{ij}\right)
\left({\bf s}_j \cdot \hat{\bf r}_{ij}\right)\right]_{\rm av.}$ due to
equal contribution from ${\bf s}_i$ and ${\bf s}_j$
parallel to $\hat{\bf r}$, and perpendicular to $\hat{\bf r}$.
As the dynamics proceeds further, the increased couplings could
well freeze ${\bf s}_i$ to a particular pattern.
The interactions then follow suit, and become correlated to the frozen orientations.
Such a scenario could well account for the correlations between OP and connectivity
observed in the tree shrew\cite{bio:shrew}.
Other procedures for obtaining synaptic couplings from input activities
\cite{Shouval,Linsker86,Bialek}, once generalized to orientations with proper
correlations, lead to similar results.
However, our intention is not to promote a particular scenario, but to
emphasize that any interactions not specifically ruled out by symmetry will 
generically be present. 
In the following, we shall explore some consequences
of joint rotation symmetry on evolution of the patterns.

\subsection{Linear analysis}\label{linear}

To underscore the difference between the two forms of rotation symmetry, let us consider the
regime of {\em linear evolution} which is analytically tractable.
Due to translation symmetry, the problem is simplified in terms of
the Fourier modes 
${\tilde s}_\alpha\left({\bf q},t\right)
=\int d^2 x e^{i{\bf q\cdot x}}s_\alpha\left({\bf x},t\right)$,
where $\alpha=1,2$ (or $x,y$) labels the two components of the vector ${\bf{\tilde s}}$.  
After Fourier transforming the interactions $J(R)$ and $K(R)$ introduced above, 
the linear evolution equation takes the form 
\be\label{L(q)}
\partial_t {\tilde s}_\alpha\left({\bf q},t\right)= \sum_{\beta=1,2} \left[J(q)\delta_{\alpha\beta}
+q_\alpha q_\beta K(q)\right]{\tilde s}_\beta\left({\bf q},t\right).  
\ee 
Due to the assumed isotropy, the functions $J$ and $K$ only 
depend on the magnitude of the vector $\bf q$.  
For example, they can be band-pass filters peaked at $\overline q=2\pi/\Lambda$,
to reproduce the power spectrum of cortical maps.
In the case of full rotation symmetry, invariance of the equations under independent
rotations of $\bf s$ and $\bf r$ requires $K(q)=0$. 
However, if $\bf s$ and $\bf r$ can only be rotated together, 
a finite $K(q)$ is possible and should be generically present.
(One way to see this is that ${\bf q}\cdot{\tilde {\bf s}}$ 
is invariant under joint rotations, 
but not  separate rotations of $\tilde{\bf s}$ and ${\bf r}$.)

A finite $K(q)$ mixes the evolution of the two components ${\tilde {\bf s}}_1$
and ${\tilde {\bf s}}_2$.
This mixing can be removed 
by decomposing the field ${\tilde {\bf s}}$ into 
{\em longitudinal} and {\em transverse} components.
For a given ${\bf q}$, the longitudinal component is parallel to $\bf q$,
and the transverse component is perpendicular to it.
Under the action of the linear operator in Eq.~(\ref{L(q)}), the two components 
grow as $e^{\left[J(q)+q^2K(q)\right]t}$ and $e^{J(q)t}$. 
If $K(q)=0$ (full rotation symmetry) the two modes grow at the same rate,
over a time scale $\tau_1(q)\sim 1/J(q)$. 
Even a small $K(q)$ breaks this degeneracy, introducing a second time scale
$\tau_2(q)\sim 1/[q^2K(q)]$ over which the effects of anisotropy become apparent.

\begin{figure}
\centering
\includegraphics[height=4cm,width=7cm]{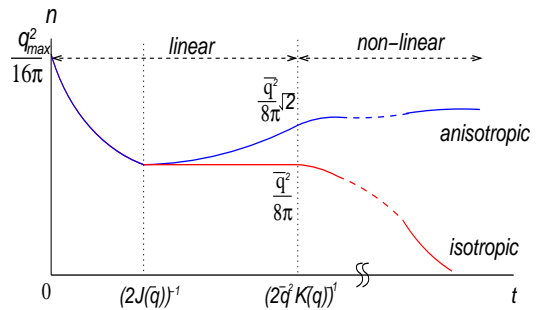}
\caption[]{Schematic depiction of the evolution of the density of zeros
 for isotropic(dark line) and anisotropic (gray line) interactions.
Anisotropy results in an increase of the density of pinwheels
in the latter stages of linear regime.
The non-linear extrapolation is based on simulation results.
}
\label{fig:dynamics}
\end{figure}

Note that when the two modes grow at the same rate ($K(q)=0$),
an equal superposition to these modes is compatible with 
a rainbow pattern which does not contain any nodes.
(Of course the rainbow is one of many possible patterns.)
However, $K(q)$ is generically non-zero for a joint rotation
symmetry, and one of the two modes eventually dominates the other.
The dominance of transverse or longitudinal components increases the
density of zeros, and is incompatible with rainbow patterns.
We repeated the analysis of WG for the density of pinwheels in the linear
regime, in the presence of a small $K(q)$.
The calculation is cumbersome and relegated to the supporting information,
but the final result for the evolution of pinwheel density is depicted in Fig.~\ref{fig:dynamics}.
The initial random pattern has a high density which rapidly decrease in a time of order 
$\tau_1\sim J\left(\overline q\right)^{-1}$ to the limiting value of
$\pi/\Lambda^2$ predicted by WG.
This is the case for
both isotropic$\left(K(q)=0\right)$ and anisotropic $\left(K(q)\neq 0\right)$ cases.
However, pinwheel density then goes up by a
factor of approximately $\sqrt{2}$ for the anisotropic case
on a time scale of $\tau_2\sim \left[\overline{q}^{2}K\left(\overline q\right)\right]^{-1}$
while it remains as $\pi/\Lambda^2$ for the isotropic case.
As explained in the supporting information the factor of $\sqrt{2}$ is the outcome of an approximate
evaluation of density which is analytically tractable.
We also performed simulations that confirmed an increase in density by a small 
factor of $\sim$ 1.12. While the increase in density is small, nonetheless
implies (pair) creation of pinwheels in the anisotropic case,
a phenomenon that is absent in the isotropic models.
Note that the ultimate density ratio between isotropic and anisotropic cases is
a universal number, independent of the degree of anisotropy.
The strength of $K(q)$ only dictates the time scale over which the density
increases, and not its ultimate value.

\subsection{Simulations}\label{simulation}
While the above arguments from the linear regime strongly suggest
that joint rotational symmetry promotes pinwheel stability, 
verification of this hypothesis comes from simulations of the 
{\em nonlinear} evolution.
For the latter, $\{{\bf s}_i(t)\}$ was placed on a lattice of points 
of locations ${\bf r}_i$, and evolved in time according to  
\begin{eqnarray}\label{discrete}
\partial_t {\bf s}_i & = & {\bf s}_i\left(1-\mid {\bf s}_i\mid^2\right) \nonumber \\
 & + & \sum_j\left[J\left(r_{ij}\right)~{\bf s}_j +
K\left(r_{ij}\right)~\left({\bf s}_j\cdot \hat{\bf r}_{ij}\right) \hat{\bf r}_{ij} \right],
\end{eqnarray}
where ${\bf r}_{ij}={\bf r}_i-{\bf r}_j$ has magnitude $r_{ij}$ 
along the unit vector $\hat{\bf r}_{ij}$.
The nonlinearity appearing in the first term on the right hand side
stabilizes the magnitude of ${\bf s}_i$ to unity.
The linear evolution is governed by
a {\em vectorial center--surround filter}, composed of two parts:
{\em (a)} A standard center--surround filter with positive couplings
$J_s$ in a circle of size $R/2\sim \Lambda$ 
and negative values $J_l$ in an annulus from $R/2$ to $R$.
{\em (b)} Additional couplings in the annular region that explicitly depend
on orientations relative to the lines joining lattice points,
and invariant only under joint rotations. 
We employ positive long--range couplings $K$, 
to mimic the preferential `horizontal' connectivity of co-oriented 
co-axially aligned receptive fields, as  reported in Ref.~\cite{bio:shrew}.
(Similar kinds of anisotropic interactions were also employed 
in a model for dynamics of neural activity in the visual cortex\cite{Bressloff}.
The anisotropic coupling by lateral neural connectivities was also obtained and
associated with pinwheel structure in Ref.\cite{Linsker86}.)

\begin{figure}
\centering
\includegraphics[height=5cm,width=5cm]{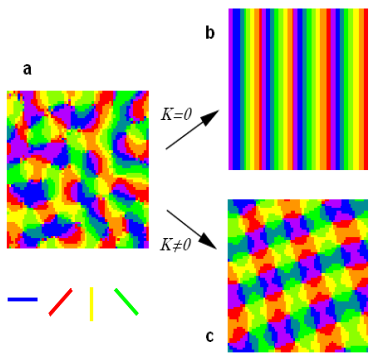}
\caption[]{{\bf a} The development of a random initial condition by a typical
center--surround (bandpass) filter leads to a collection of pinwheels.
The filter used in {\bf b} has full rotation symmetry ($K(r)=0$ in Eq.~(\ref{discrete})).
In this case the pinwheels annihilate in pairs, giving way to a rainbow pattern at long times.
{\bf c} By contrast, a model with joint rotation symmetry evolves to a stable pattern
of pinwheels. This figure was generated by the {\em vectorial center--surround} filter in
Eq.~(\ref{discrete}), with a non-zero $K(r)$.}
\label{fig:model}
\end{figure}

Simulations are started on an $L \times L$ lattice with
initial values of $|{\bf s}_i|=10^{-3}$, equally distributed over
all angles, with $J_s=0.01$, $J_l=-0.0039$, and $R=10$.
As shown in Fig.~\ref{fig:model}a, undifferentiated initial conditions
quickly develop into a pattern with pinwheels reminiscent of actual maps.
Further evolution depends on the symmetry of development rules.
Full rotation symmetry with $K=0$, and the action of (b) above turned off,
leads to a rainbow state with no pinwheels at long times, 
as in Fig.~\ref{fig:model}b.
However, reduction of this symmetry by adding interactions in (b) with $K=0.0039$, 
above eventually results in a square lattice of pinwheels, as in Fig.~\ref{fig:model}c.
Naturally, we do not imply that pinwheels in cortical maps form a square lattice
(various inhomogeneities could easily trap these vortices in a distorted arrangement),
but that they are intrinsically stable under such development rules.
The precise choice of long--range couplings is not important in this regard, 
and we observed pinwheel patterns with other types of anisotropic coupling 
(some also available as Fig.~\ref{fig:negative-K} in the supporting information).

\begin{figure}
\centering
\includegraphics[height=3cm,width=8cm]{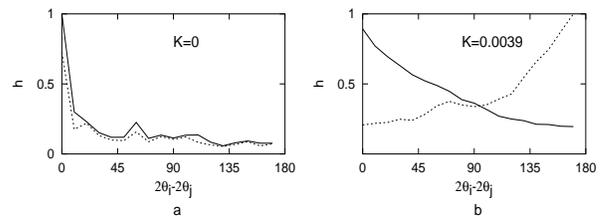}
\caption[]{ 
Histograms of relative angles for $2(\phi_{ij}-\theta_j)=0^o$(solid line)
and $2(\phi_{ij}-\theta_j)=90^o$(dotted line)
with isotropic(K=0) (a) and anisotropic(K=0.0039) (b) interactions.}
\label{fig:modelhisto}
\end{figure}

\begin{figure}[h]
\centering
\includegraphics[height=9cm,width=10cm]{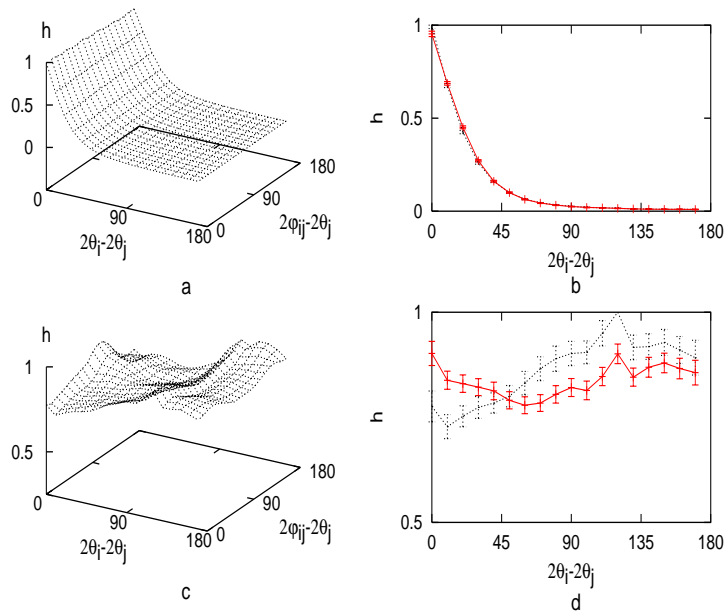}
\caption[]{
Histograms of OP from a map of cat.
Here $\theta_i(\theta_j)$ is the preferred orientation at a pixel $i(j)$,
and $\phi_{ij}$ is the angle of the line joining points $i$ and $j$.
The left column displays full histograms 
as functions of relative orientations,
as well as orientations relative to the line joining the
two pixels $i$ and $j$. 
In the right column,
solid lines represent histograms for $2(\phi_{ij}-\theta_j)=0^{o}$,  
and dotted lines represent histograms for $2(\phi_{ij}-\theta_j)=90^{o}$.  
${\bf a}$ and ${\bf b}$ correspond to separations of 5-10 pixel spacings
and ${\bf c}$ and ${\bf d}$, to separations of 55-60 pixels.
}
\label{fig:cathisto}
\end{figure}

\begin{figure}[h]
\centering
\includegraphics[height=4cm,width=4cm]{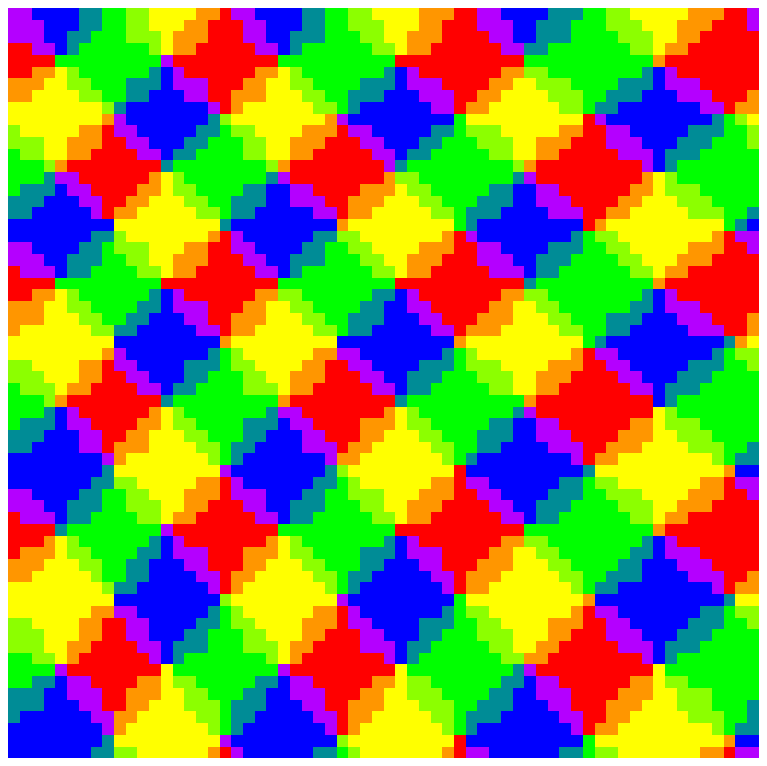}
\includegraphics[height=4cm,width=4cm]{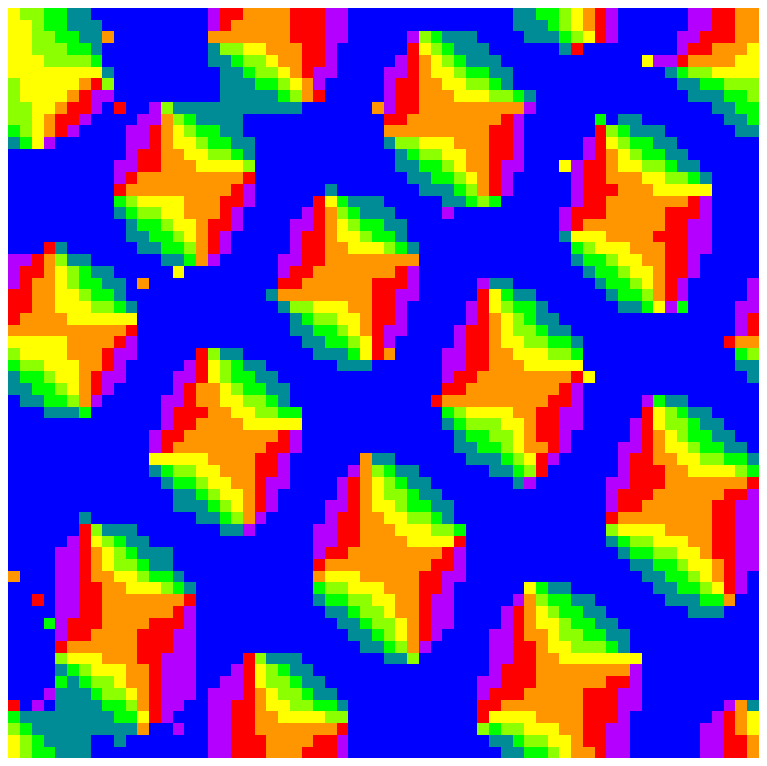}
\vspace*{0.1cm}
\\a \hspace*{3.cm} b\\
\caption[]{{\bf a.} Stable pattern of pinwheels with joint rotation symmetry,
with a negative value of $K=-0.0039$. (Compare with
Fig. 4c of the manuscript.)
{\bf b.} A pinwheel pattern is also generated in simulations where the
angles are constrained to the region of [0,$\pi$]. 
}
\label{fig:negative-K}
\end{figure}

Not surprisingly,  the anisotropic couplings lead to correlations
between OP and the topographic angles.
We repeated the histogram analysis of actual maps with those generated by
numerical simulations, and some results are plotted in Fig.~\ref{fig:modelhisto}.
There is no dependence on topography for $K=0$,
as depicted in Figure.~\ref{fig:modelhisto}a which shows
two relative angle histograms for $2 \left( \phi_{ij}-\theta_j \right )=0^o$ and
$2 \left( \phi_{ij}-\theta_j \right)=90^o$.
For $K\neq 0$, there are positive correlations in relative angles
for $2 \left(\phi_{ij}-\theta_j \right)=0^o$
and negative correlations for
$2 \left(\phi_{ij}-\theta_j \right)=90^o$[Fig.~\ref{fig:modelhisto}b].
The topographic dependence is robust , and does  not significantly depend on
the strength of the anisotropic coupling.

\section{Conclusions}\label{conclusions}

Collinearity is a prominent characteristic of line segments in natural images. 
It is reasonable to expect that cortical maps of OP reflect a corresponding tendency.
A basic consequence of the tendency of line segments to be collinear is 
the absence of a full rotation symmetry, independent of the underlying topography.
We demonstrate the lack of full symmetry by analyzing histograms of monkey and
cat maps.
We then explore consequences of reduced symmetry on the behavior of
evolving fields of OP.
In the linear regime, we find that the new interactions allowed generates a new time
scale over which the pinwheel density can actually increase.
Numerical simulations confirm that this tendency persists in the non-linear regime,
resulting in patterns with stable pinwheels.

While the stability problem of pinwheels in OP maps is not widely appreciated, 
it has been the motivation for two other recent studies. 
In Ref.~\cite{KC} a different  coupling between neurons is used based on 
a `wiring length minimization' principle,
while in Ref.~\cite{Wthesis} higher order non-linearities are employed in place of the
stabilizing ${\bf s_i}\mid {\bf s}_i\mid^2$ term in Eq.~(\ref{discrete}).
While these models lead to stable patterns of pinwheels, they can not account for the 
`anisotropic' features of actual OP maps, as both have full rotation symmetry.
A potential relation between the symmetries and correlations of line segments 
in natural images, and the statistics of OP maps (including stability and arrangement of pinwheels),
may provide further clues to how visual information is processed by the brain.

\noindent This work was supported by the NSF grant number DMR-01-18213.
We are grateful to F. Wolf, D. Chklovskii, 
for illuminating discussions, and to M. Sur and K. Obermeyer
for sharing of their data.

\narrowtext 


\widetext

\begin{center}
{\bf {SUPPORTING INFORMATION}}
\end{center}

{\noindent \bf Density of Zeros for a Field with Anisotropy}

\noindent We consider a field $\vec{S} \equiv (S_x(x,y), S_y(x,y))$.
Subject to joint rotational invariance of $\vec{S}$ and $\vec{r} \equiv (x,y)$,
the most general Gaussian weight is
\be  
p\left[\vec{S}\,\right]\propto  
\exp  \left \{ -{1 \over 2} \int {{d^2 q} \over {(2\pi)^2}} 
S_{\alpha}(\vec{q}\,) S_{\beta}(-\vec{q}\,) 
 \left [ 
\lambda^{-1}(q) {{q_{\alpha} q_{\beta}} \over {q^2}} 
+\tau^{-1} (q) \left( \delta_{\alpha \beta}
-{{q_{\alpha} q_{\beta}} \over {q^2}} \right) 
\right ]   \right \},
\ee
where $\vec{S}(\vec{q}\,)$ is the Fourier transform of $\vec{S}(\vec{r}\,)$,
$\lambda (q)$ and $\tau(q)$ are longitudinal and transverse contents of the power
spectrum, and $\tau(q)=\lambda(q)$ in an isotropic system.
The average density of zeros is obtained from\cite{Halperin}
\be
n=  \left < {\delta}^2 (\vec{S}\,)~ {\rm det}~ \partial_{\alpha} S_{\beta} 
\right >
=\left < {\delta}^2 \left(\vec{S}(0)\right) \right >  \left < \left | 
\partial_x S_x(0) \partial_y S_y(0) -\partial_x S_y(0) \partial_y S_x (0)
\right | \right >.
\label{eq:avdensity}
\ee
Here the average can be taken at $\vec{r}=0$ because of translation symmetry.
Also, the averages $\partial_\alpha\vec{S}$ are independent, 
because the probability
distribution function is invariant
under $\vec{S} \rightarrow \vec{S} + \vec{C}$.
The first average is easily calculated as
\be
\left < {\delta}^2 \left(\vec{S}(0)\right) \right >= \int {{d^2 k} \over {(2 \pi)^2}} 
\left < e^{i\vec{k} \cdot \vec{S}(0)} \right >
= \int {{d^2 k} \over {(2 \pi)^2}} 
\exp \left [ -{k^2 \over 2} \left < \vec{S}(0) \cdot \vec{S}
(0) \right > \right ].
\label{eq:deltas}
\ee
The variance of $\vec{S}(0)$ is
\be
\left < \vec{S}(0) \cdot \vec{S} (0) \right >
= \int {{d^2 q d^2 q'} \over {(2 \pi)}^4}  
\left < S_{\alpha}(\vec{q}\,) S_{\alpha}(\vec{q}\,') \right >
=\int {{d^2 q d^2 q'} \over {(2 \pi)}^4} (2 \pi)^2 \delta^2(\vec{q}+\vec{q}\,')
~\left[\lambda(q)+\tau(q)\right],
\label{eq:s0s0}
\ee
and by inserting Eq.~(\ref{eq:s0s0}) to Eq.~(\ref{eq:deltas}), we get
\be
\left < {\delta}^2 \left(\vec{S}\,\right) \right >
={1 \over {2 \pi}} {1 \over {\int {{d^2 q} \over (2 \pi)^2}
\left[\lambda(q)+\tau(q)\right]}}=
{1\over\int dq~q\left[\lambda(q)+\tau(q)\right]}.
\label{eq:d2s}
\ee

As a first step to calculating the average determinant, we consider
\ber
\left < \partial_i S_{\alpha} (0) \partial_j S_{\beta} (0)  \right >
&=& \int {{d^2 q} \over {(2 \pi)^2}} {{d^2 q'} \over {(2 \pi)^2}}
(i q_i) (i q'_j) \left < S_{\alpha} (\vec{q}\,) S_{\beta} (\vec{q}\,') \right > \nonumber \\
&=& \int {{d^2 q} \over {(2 \pi)}^2} q_i q_j \left [ 
\lambda(q) {{q_{\alpha} q_{\beta}} \over q^2} 
+\tau(q) \left(\delta_{\alpha \beta}
-{{q_{\alpha} q_{\beta}} \over q^2}\right) 
\right ]
\nonumber \\
&=& \int {{d^2 q} \over {(2 \pi)^2}} q^2  \left [ 
\tau(q) 
\left ( {{\delta_{ij} \delta_{\alpha \beta}} \over 2}
-{{\delta_{ij} \delta_{\alpha \beta}+\delta_{i \alpha} \delta_{j \beta} 
+ \delta_{i \beta} \delta_{j \alpha}} \over 8} \right ) \right .
\nonumber \\
&& \hspace*{2cm} + \left . \lambda(q) {{\delta_{ij} \delta_{\alpha \beta}+\delta_{i \alpha} \delta_{j \beta} 
+ \delta_{i \beta} \delta_{j \alpha}} \over 8} \right ]
\nonumber \\
&=&{{\delta_{ij} \delta_{\alpha \beta}} \over 4 \pi} \int dq q^3 \tau(q)
+{\delta_{ij} \delta_{\alpha \beta}+\delta_{i \alpha} \delta_{j \beta}
+ \delta_{i \beta} \delta_{j \alpha} \over {16 \pi}}
\int dq q^3 \left(\lambda(q)-\tau(q)\right).
\label{eq:rs}
\eer
We next rewrite Eq.~(\ref{eq:rs}) as
\be\label{u-corr}
\left < \partial_i S_{\alpha} \partial_j S_{\beta}  \right >
= {\delta_{ij} \delta_{\alpha \beta}}~ \kappa 
+ (\delta_{ij} \delta_{\alpha \beta}+\delta_{i \alpha} \delta_{j \beta}
+ \delta_{i \beta} \delta_{j \alpha})~ \mu ,
\ee
where $\kappa= \int dq~ q^3~ \tau(q)/(4\pi)$, and
$\mu= \int dq~ q^3~ (\lambda(q)-\tau(q))/(16\pi)$ is zero in an isotropic system.
In the isotropic system, each of the four derivatives is an independent variable.
However, for $\mu\neq0$, there are two correlated pairs $\left(\partial_x S_x,\partial_y S_y\right)$,
and $\left(\partial_x S_y,\partial_y S_x\right)$. 
For the first pair, we have $\left<\left(\partial_x S_x\right)^2\right>=
\left<\left(\partial_y S_y\right)^2\right>=\kappa+3\mu$, whereas for the
second pair, $\left<\left(\partial_y S_x\right)^2\right>=
\left<\left(\partial_x S_y\right)^2\right>=\kappa+\mu$.
The cross correlations in each pair are identical,
$\left<\partial_x S_x\partial_y S_y\right>=\left<\partial_x S_y\partial_y S_x\right>=\mu$,
such that the average value of the determinant is zero.

As a second step toward the calculation of average absolute value of the
determinant, we find its probability distribution as 
\be 
p(d)= \left < \delta \left[d - \left(\partial_x S_x \partial_y S_y -\partial_x S_y \partial_y S_x 
\right)\right] \right >
= \int {{d\omega } \over {2 \pi}} e^{i\omega d} \left < 
e^{i\omega  (\partial_x S_x \partial_y S_y -\partial_x S_y \partial_y S_x )} \right > .
\label{eq:pdfd}
\ee
As established above, the two factors in the final exponent
are independent random elements. 
The random variables $\partial_\alpha S_\beta\equiv u_{\alpha\beta}$ are Gaussian
distributed, with covariances given by Eq.~(\ref{u-corr}).
By inverting the covariance matrix, we can construct the probability distribution
for $\{u_{\alpha\beta}\}$ and then calculate the average
\ber
\left < e^{i\omega  u_{xx}u_{yy}} \right > 
&=& \int {du_{xx} du_{yy} \over {\cal N}}~\exp \left \{ -{1 \over 2} (u_{xx}, u_{yy})
\left [ 
\begin{array}{cc}
{\kappa+\mu \over (\kappa+\mu)^2-\mu^2} & - {\mu \over (\kappa+\mu)^2-\mu^2}-i\omega  \\
{-{\mu \over (\kappa+\mu)^2-\mu^2}-i\omega } & {\mu \over (\kappa+\mu)^2-\mu^2}
\end{array} 
\right ]
\left ( 
\begin{array}{c}
u_{xx} \\
u_{yy}
\end{array}
\right )
\right \}
\nonumber \\
&=&\left [ {(\kappa+\mu)^2 \over ((\kappa+\mu)^2-\mu^2)^2}
-{\mu^2 \over {((\kappa+\mu)^2-\mu^2)^2}}
-{2i\omega \mu \over {(\kappa+\mu)^2-\mu^2}} 
+ \omega ^2 \right ]^{- {1 \over 2}}
\nonumber \\
&& \times \left [ { (\kappa+\mu)^2-\mu^2 \over 
{((\kappa+\mu)^2-\mu^2)^2}} \right ]^{1 \over 2}
\nonumber \\
&=& \left [ 1+\omega ^2 ((\kappa+\mu)^2 -\mu^2) - 2 i \omega  \mu \right ]^{-{1 \over 2}}.
\label{eq:xx}
\eer
(The normalization ${\cal N}$ in the denominator is simply the numerator evaluated at $\omega=0$.)
Similarly, the second average is
\be 
\left < e^{-i\omega  u_{xy}u_{yx}} \right >
=\left [ 1+\omega ^2 ((\kappa+3 \mu)^2 -\mu^2) + 2 i \omega  \mu \right ]^{-{1 \over 2}}.
\label{eq:xy}
\ee
Inserting Eqs.~(\ref{eq:xx}) and (\ref{eq:xy}) into Eq.~(\ref{eq:pdfd}) gives the implicit result
\be \label{pdf}
p(d)=\int_{-\infty}^{\infty}
{d\omega  \over 2 \pi}
{e^{i\omega d} \over 
[1-2i\omega \mu+\omega ^2(\kappa^2+ 2 \mu \kappa)]^{1 \over 2}
[1+2i\omega \mu+\omega ^2(\kappa^2+ 6 \mu \kappa+8 \mu^2)]^{1 \over 2} }.
\ee 

Let us consider the isotropic case, $\mu=0$.
The probability distribution function is
\be
p(d)=\int_{-\infty}^{\infty} {d\omega  \over 2 \pi} {e^{i \omega  d} \over {1+\kappa^2 \omega ^2}}
= {1 \over 2 \kappa} e^{{-|d|} / \kappa},
\ee
from which we obtain
\be
\left< |d| \right> =2 \int_0^{\infty} 
d x {1 \over {2 \kappa}} e^{-x / \kappa}=\kappa.
\label{eq:avd}
\ee 
From Eqs.~(\ref{eq:avdensity}), (\ref{eq:d2s}), and (\ref{eq:avd}),
average density of pinwheels is then
\be
n={1 \over {4 \pi}} {{\int dq q^3 \tau(q)} \over 
{\int dq q (\lambda(q)+\tau(q))}}={1 \over {8 \pi}} {{\int dq q^3P(q)} \over 
{\int dq q P(q)}},
\ee
where $P(q)=\lambda(q)+\tau(q)=2\tau(q)$ is the power spectrum of the field.
The above result is smaller by a factor of two than that 
obtained in ref.~\cite{WG}.
However, our calculation was with a vector field, 
whereas the orientation preference
is a director field, which is the same if the vector is inverted.
To incorporate this feature, ref.~\cite{WG} works 
with a complex field $|z(\vec{x}\,|
e^{2i\theta(\vec{x}\,)}\equiv (S_x+iS_y)^2$,
a procedure that doubles the zeros calculated above for the field $(S_x+iS_y)$.
This factor is not important to us, because 
we are interested in how the result is
modified by anisotropy.

Performing the integral in Eq.~(\ref{pdf}) for $\mu\neq0$ is not an easy task.
We note that because $\left<d\right>=0$, the average of the absolute value provides
a measure of the width of the probability distribution $p(d)$.
A similar measure of the width of the distribution that is much easier to calculate
is the standard deviation  $\sqrt{<d^2>}$.
Using standard properties of Gaussian distributed variables, the variance of
$d$ is calculated as
\ber
\left<d^2\right>&=&\left<\left(\partial_x S_x \partial_y S_y 
- \partial_x S_y \partial_y S_x\right)^2\right>
\nonumber \\
&=&\left<\left(\partial_x S_x\right)^2\right>\left<\left(\partial_y S_y\right)^2\right>
+2\left<\partial_x S_x \partial_y S_y\right>^2+\left<\left(\partial_x S_y\right)^2\right>
\left<\left(\partial_y S_x\right)^2\right>
\nonumber \\
&&\hspace*{0.2cm}+2\left<\partial_x S_y \partial_y S_x\right>^2
-2\left<\partial_x S_x \partial_y S_y\right>\left<\partial_x S_y \partial_y S_x\right>
\nonumber \\
&=&(\kappa+3 \mu)^2+2 {\mu}^2 + (\kappa+\mu)^2+ 2 {\mu}^2 -2 {\mu}^2
\nonumber \\
&=&2 \kappa^2 + 8 \mu \kappa + 12 \mu^2.
\eer
As measures of the width of the distribution, $\left<|d|\right>$ and
$\sqrt{\left<d^2\right>}$ should vary together.
For our estimate, we shall assume that they are proportional and choose
a proportionality constant that makes the two expressions equal for $\mu=0$;
i.e., we make the replacement
\be
\left<|d|\right>\to\sqrt{\left<d^2\right> \over 2}
=\sqrt{\kappa^2 + 4 \mu \kappa + 6 \mu^2},
\ee
resulting in the density of zeros
\be
n \approx {\sqrt{\kappa^2 + 4 \mu \kappa + 6 \mu^2} 
\over \int dq~q~(\lambda(q)+\tau(q))}.
\label{eq:n}
\ee
Using the expressions for $\kappa$ and $\mu$, we note that
\be
\kappa+2 \mu= {1 \over 4 \pi} \int dq ~q^3~ \left (\tau(q)
+{\lambda(q)-\tau(q) \over 2} \right )={1 \over 8 \pi} \int dq q^3 P(q),
\label{eq:k2m}
\ee
where $P(q)\equiv \lambda(q)+\tau(q)$ is the total power content at $q$.
With the aid of Eq.~(\ref{eq:k2m}), Eq.~(\ref{eq:n}) now becomes
\be
n \approx {\sqrt{(\kappa+2\mu)^2+(2\mu)^2} \over \int dq~q~P(q)}
={1 \over 8 \pi} {\int dq~q^3~P(q) \over \int dq~q~P(q)}
\sqrt{1+\left [ {\int dq~q^3~(\lambda(q)-\tau(q)) \over \int dq~q^3~
(\lambda(q)+\tau(q))} \right ]^2}.
\ee
For a fixed $P(q)$, the density of zeros is minimum in the isotropic
limit of $\tau(q)=\lambda(q)$.
In the extreme anisotropic limit of $\tau(q)=0$ or $\lambda(q)=0$,
the density of zeros increases by a factor of $\sqrt{2}$.
In view of the approximations involved, we also performed numerical
simulations to check whether the density of pinwheels is higher
in the anisotropic case. We found that this is indeed the case,
although the relative increase in density of 1.12 is less than the value
of $\sqrt{2}$.

 Let us illustrate the time evolution of the density of zeros, using a simple
linear model for development of the field, in which the longitudinal and
transverse components of the power spectrum grow as
\ber
\lambda(q,t)&=&\lambda_0(q)~e^{r_l(q)~t} ,\nonumber \\
\tau(q,t)&=&\tau_0(q)~e^{r_t(q)~t} ,
\eer
where growth rates are $r_l(q)=2[J(q)+q^2K(q)]$ and $r_t(q)=2J(q)$.
If initially $\lambda_0(q)=\tau_0(q)={P_0/2}$, for $q<q_{\rm max}$, 
i.e., an isotropically random initial condition, the density of zeros starts as
\be
n(t=0) \approx {1 \over 16 \pi} q_{\rm max}^2.
\ee
As time goes on, modes with the largest growth rate dominate, reducing
$n$ through pair annihilations.
Assuming small anisotropy, such that $r_l(q) \approx r_t(q) = 2 J(q)$ 
with a maximum
at $\bar{q}={2 \pi / \Lambda}$, we have
\be 
n\left(t \geq {(2J(\bar{q}))}^{-1} \right) \approx {1 \over 8 \pi} {\bar q}^2
={\pi \over 2 {\Lambda}^2}.
\ee
However, because of small anisotropy $(r_l(q) \neq r_t(q))$,
one of these nearly degenerate modes will dominate
the other, such that for longer times,
\be
n\left(t \geq {(2 {\bar{q}}^2 K(\bar{q}))}^{-1} \right)
\approx {1 \over 8 \pi} {\bar q}^2 \sqrt{2}
={\pi \over 2 {\Lambda}^2} \sqrt{2}.
\ee
Fig. 3 shows schematic evolution of $n$ for isotropic and
anisotropic cases, the increase of the density in the latter must also involve
creation of pairs of vortices.

\noindent {\bf Cortical Map of Cat}
\\
\noindent We also measured joint histograms, 
$h_R[2(\theta_i-\theta_j),2(\phi_{j}-\theta_j)]$,
for the map of cat in a manner similar to the monkey.
The size of the cat map is 204$\times$372 pixels, each
representing a region of linear size 13 $\mu$m.
As in the case of monkey map, 
we display histograms for short separations of 5-10 pixel spacings 
[Figs.~6a and b]
and separations comparable to pinwheel separations of
55-60 [Figs.~6c and d].
There is no dependence on the relative angle for short distances, 
but such a dependence appears on distances comparable to pinwheel separations. 
This again indicates a lack of full rotation symmetry in
the map of cat.
To estimate errors, we average $>$ 2,000 histograms, 
each of which is constructed
by random samplings with 2.9\% pixels of the cat map. 

\newpage

\noindent {\bf Pinwheel Patterns with Various Anisotropic Couplings}
\\
\noindent Because we do not claim to know the precise form of interactions that lead to
cortical patterns, we should at least show that our conclusions are not sensitive
to specific choice of interactions.
We tested a variety of long-range interactions in our numerical simulations
and found that pinwheels are generally present in the presence of anisotropy.
As an example, 
we observe a pinwheel pattern with a negative value of 
$K$, which is used to generate
the map in Fig.~7a.
Another potential concern is that in our simulations, the orientations are
represented by a vector, whereas in actuality they should be modeled
by a director field (vectors without arrows).
We also performed simulations in which all angles were explicitly limited to the
range from 0 to $\pi$.
Fig. 7b displays the result of such a simulation,
once more resulting in a pinwheel pattern (for interaction strength of $K=0.0039$).  

The type of anisotropy introduced above, which couples rotations of orientation
and topography, should not be confused with the anisotropy corresponding
to preference for a particular direction.
In fact, we find that both maps of monkey and cat show 
a predominance of certain orientations.
The histogram of orientations for monkey is shown
in Fig.~8a and for cat, in Fig.~8b.

\begin{figure}[h]
\centering
\includegraphics[height=6cm,width=12cm]{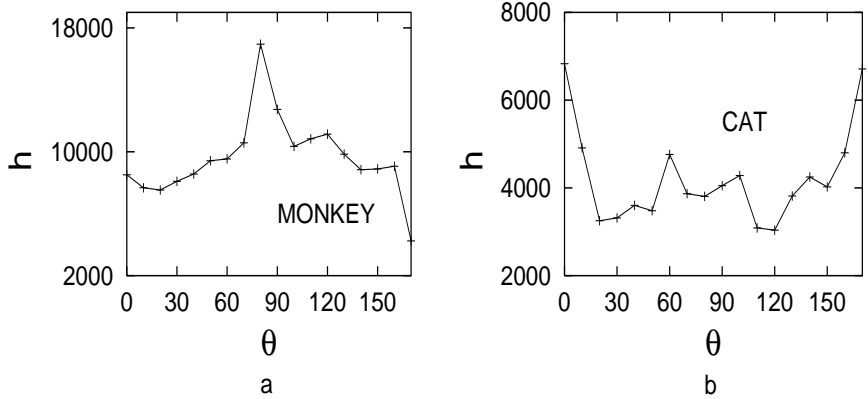}
\caption[]{Predominance of certain orientation in the maps of monkey (a) and
cat (b).
All orientations are not equally present in either map.}
\label{fig:histopredomi}
\end{figure}

We test the possibility of stabilizing pinwheels with this form
of anisotropy by numerical simulations in models with
a preference for the horizontal direction,
using the time evolution
\be
\label{eq:magnetic}
\partial_t {\vec s}_i  =  {\vec s}_i\left(1-\mid {\vec s}_i\mid^2\right)
  +  \sum_j\left[J\left(r_{ij}\right)~{\vec s}_j +
K\left(r_{ij}\right)~\left({\vec s}_j\cdot \hat{r}_{ij}\right) \hat{r}_{ij}
\right]+\vec{H},
\ee
where $\vec{H}=H_x \hat{x}$.
Introducing the predominance of certain orientation does not
change the outcomes.
We still find that $K(r_{ij})=0$, and preference for
the horizontal direction with $H_x=0.1$, 
results in a rainbow state
with no vortices, as depicted in
Fig.~9a.
We obtain a lattice of pinwheels by adding interactions $K(r_{ij})=J_l$
with $H_x=0.1$ [Fig.~9b].
The pinwheels are thus stabilized not by the preference for a particular
angle but by the reduced symmetry of combined rotations of orientations
and the visual field.

\begin{figure}[h]
\centering
\includegraphics[height=4cm,width=4cm]{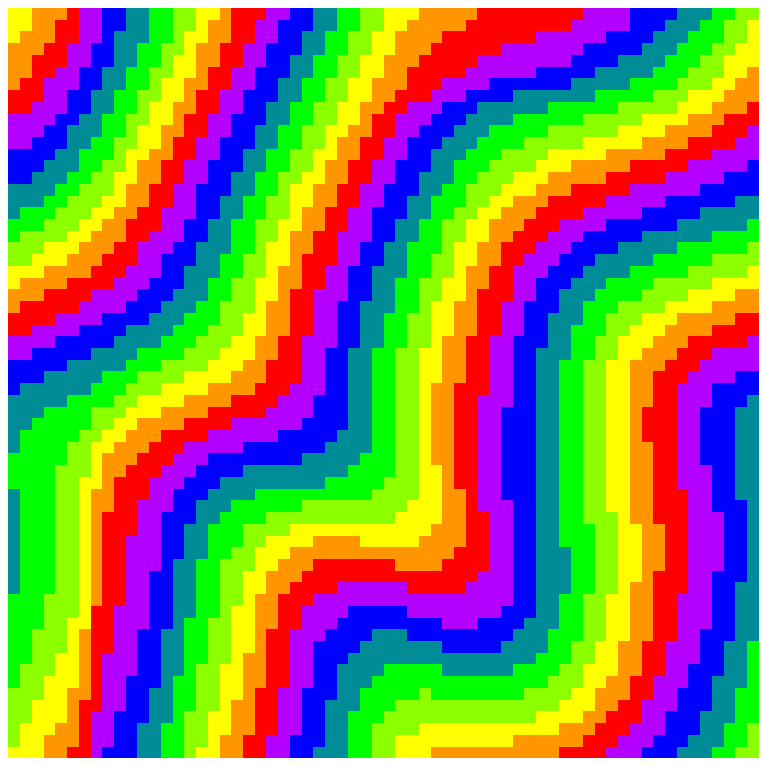}
\includegraphics[height=4cm,width=5cm]{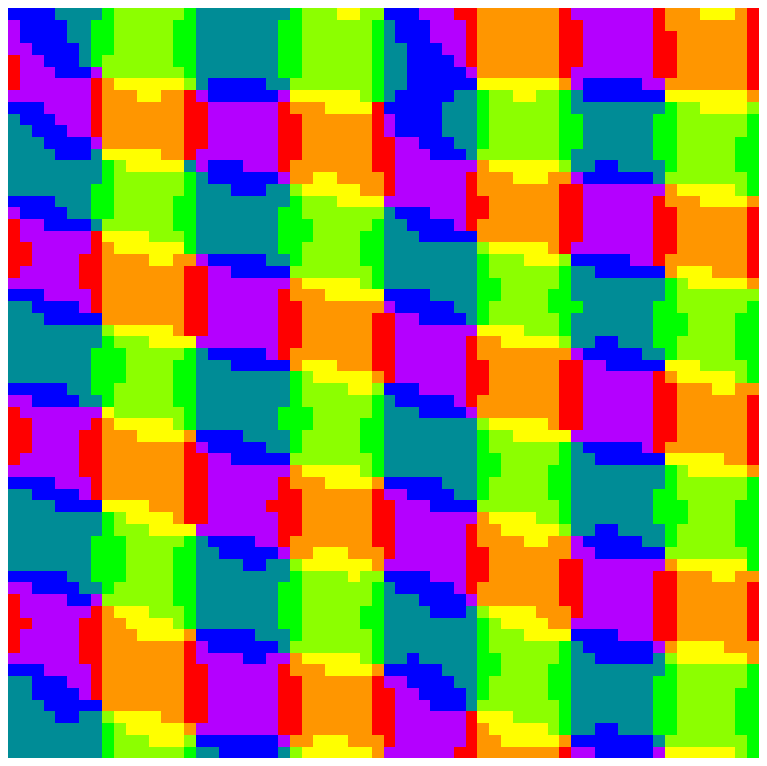}
\vspace*{0.1cm}
\\a \hspace*{3.cm} b\\
\caption[]{{\bf a.} The development of a random initial condition
by a filter with isotropic pair interactions ($K(r_{ij})=0$), 
but with a preference for
the horizontal direction ($H_x=0.1$).
Such preference does not stabilize the pinwheels. 
{\bf b.} The stabilized pinwheel pattern with anisotropic pairwise interactions
($K(r_{ij})=-0.0039$),
in addition to a preference for the horizontal direction ($H_x=0.1$).
}
\label{fig:prerainbow}
\end{figure}       

\noindent {\bf Error Estimation for Histograms of Orientation Preference}
\\
\noindent We estimate error bars in Fig.~2c and e 
by averaging $>$ 2,000 histograms.
Each histogram is calculated by random sampling
2.9 \% of total pixels in the monkey map. 
We then tested that this artificial sampling procedure does not
lead to spurious effects due to finite-size and other potential
factors by applying it to numerically generated maps.
For the latter, we generated
random pinwheel patterns through superposition
of isotropic Fourier modes having the same longitudinal
and transverse components.
To do so, we obtain the orientation at position $(i,j)$
by 2D Fourier transformation of 
$S_{\alpha}(\vec{q})(i,j)=\exp(c_1+c_2 q^2-c_3 q^4) \exp(i \Phi_{\alpha,i,j})$,
where $\alpha=x$ or $y$ and $\Phi_{\alpha,i,j}$ is a random variable
ranging from 0 to $2 \pi$.
Fig. 10 displays such an isotropic pattern
with 256$\times$256 pixels,  
with a pinwheel density close to that of the monkey map.
The angle histograms from the isotropic pattern
for $2 \left( \phi_{ij}-\theta_j \right)=0$ (red line)
and $2 \left( \phi_{ij}-\theta_j \right)=90$ (black line)
are shown in Fig.~10b.
We construct histograms for the isotropic pattern
with larger size (512$\times$512) than the monkey map(360$\times$480)
(Fig.~10c).
Here error bars in Fig.~10b and c
are estimated by averaging $>$ 2,000 histograms.
As in the analysis for the monkey map,
each histogram is constructed by randomly sampling
2.9\% of total pixels in one pattern. 
The difference between two histograms for 
$2 \left( \phi_{ij}-\theta_j \right)=0$ and   
$2 \left( \phi_{ij}-\theta_j \right)=90$ from the monkey map
is larger than the histogram differences
from both isotropic patterns with sizes 256$\times$256 and $512\times$512.
Hence we have some confidence that observed
couplings of relative orientation to the underlying
topography in the monkey map are not from statistical errors.  

\begin{figure}[h]
\centering
\includegraphics[height=4.cm,width=4cm]{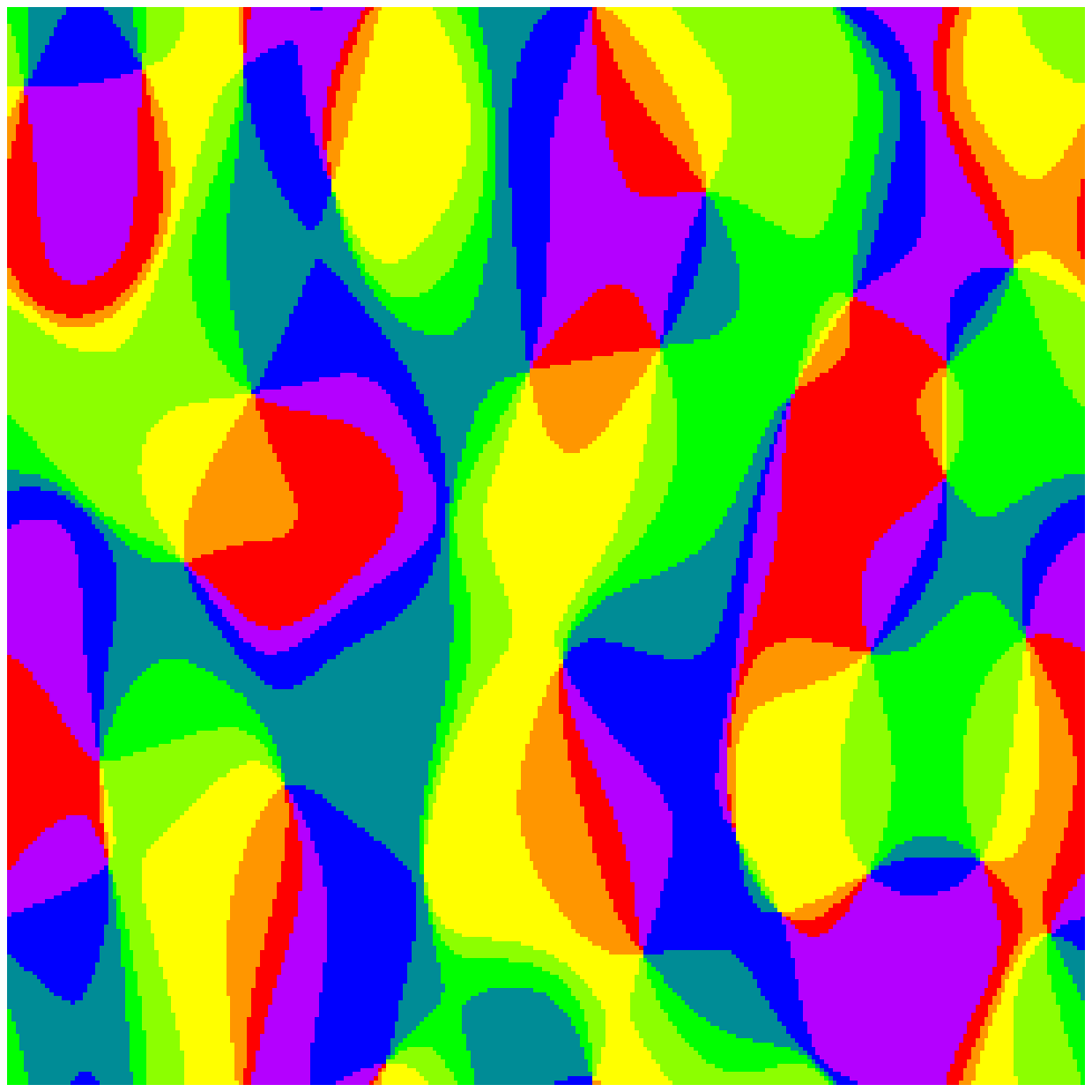}\\a\\
\includegraphics[height=5.cm,width=9cm]{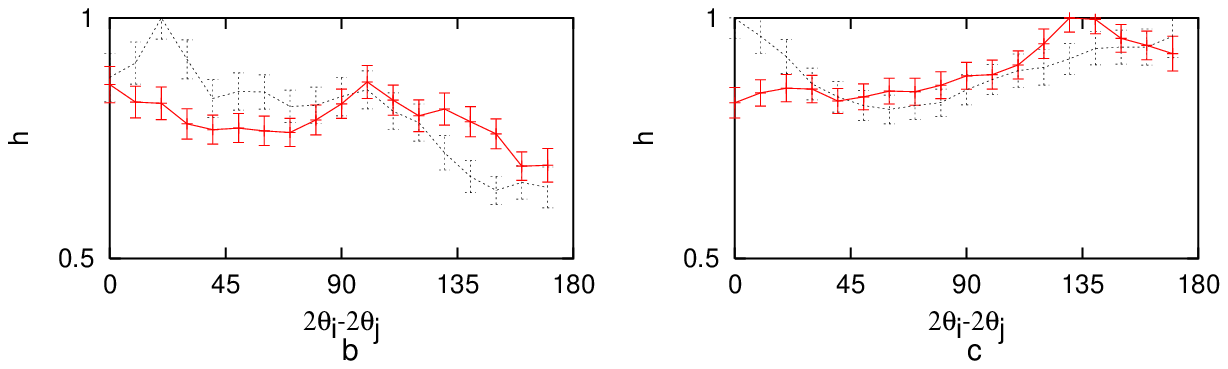}
\caption[]{
{\bf a.} A pinwheel pattern randomly generated by superposition of isotropic Fourier modes.
The size of the pattern is 256$\times$256 pixels.
The relative angle histograms for $2(\phi_{ij}-\phi_{j})=0^o$ (black)
and $2(\phi_{ij}-\phi_{j})=90^o$ (red) from patterns with 
sizes of 256$\times$256 (b) and 512$\times$512 (c).
Here error bars are estimated by averaging $>$ 2,000 histograms.
Neither pattern shows comparable topographic dependence to that in the monkey map.
}
\label{fig:isofourier}
\end{figure}




\begin{thebibliography}{7}
\bibitem{Fnoise}
Constant stirring by sufficiently strong external 
noise can also lead to dynamic
creation and annihilation of pinwheels, 
but our focus is on evolving fields where
the only randomness is in the choice of initial conditions.
\end{thebibliography}
\end{document}